\documentclass[11pt,aps,prb,preprint]{revtex4}   

\advance\oddsidemargin -.2in
\advance\evensidemargin -.2in
\advance\textwidth.4in
\usepackage{graphicx}

\usepackage{amsmath}    
\usepackage{amssymb}
\usepackage{graphicx}   
\usepackage{enumerate}

\newcommand{\calerr}{{\cal E}_{rr}}
\newcommand{\calett}{{\cal E}_{{\theta}{\theta}}}
\newcommand{\calepp}{{\cal E}_{{\phi}{\phi}}}
\newcommand{\caletp}{{\cal E}_{{\theta}{\phi}}}
\newcommand{\calert}{{\cal E}_{r{\theta}}}
\newcommand{\calerp}{{\cal E}_{r{\phi}}}
\newcommand{\calbrr}{{\cal B}_{rr}}
\newcommand{\calbtt}{{\cal B}_{{\theta}{\theta}}}

\newcommand{\calbpp}{{\cal B}_{{\phi}{\phi}}}

\newcommand{\calbtp}{{\cal B}_{{\theta}{\phi}}}

\newcommand{\calbrt}{{\cal B}_{r{\theta}}}

\newcommand{\calbrp}{{\cal B}_{r{\phi}}}

\newcommand{\cunew}{\cos{(kr\!-\!\omega t)}}

\newcommand{\sunew}{\sin{(kr\!-\!\omega t)}}
\newcommand{\calEbf}{\bs{\cal E}}
\newcommand{\calBbf}{\bs{\cal B}}

\newcommand{\EM}{E\&M}

\newcommand{\bs}[1]{\boldsymbol{#1}}

\begin{document}

\title{Comparison of electromagnetic and gravitational radiation;\\
what we can learn about each from the other}
\author{Richard H.\ Price}
 \email{richard.price@utb.edu}   
 \affiliation{University of Texas at Brownsville, 
 Department of Physics and Astronomy,
 Brownsville, TX 78520}
\author{John W.~Belcher}
\email{jbelcher@mit.edu}   
\affiliation{Massachusetts Institute of Technology,
Department of Physics, Cambridge, MA 02139}
\author{David A.~Nichols}
\email{dan87@cornell.edu
}    
\affiliation{California Institute of Technology,
Theoretical Astrophysics, Pasadena, California 91125
(Current address: 
Center for Radiophysics and Space Research, Cornell University,
Ithaca, New York 14853)
}
\date{\today}

\begin{abstract}\baselineskip=10pt
We compare the nature of electromagnetic fields and of gravitational
fields in linearized general relativity. We carry out this comparison
both mathematically and visually.  In particular the ``lines of
force'' visualizations of electromagnetism are contrasted with the
recently introduced tendex/vortex eigenline technique for visualizing
gravitational fields. Specific solutions, visualizations, and
comparisons are given for an oscillating point quadrupole
source. Among the similarities illustrated are the quasistatic nature
of the near fields, the transverse $1/r$ nature of the far fields, and
the interesting intermediate field structures connecting these two
limiting forms. Among the differences illustrated are the meaning of
field line motion, and of the flow of energy.

\end{abstract}
\maketitle

\section{Introduction}\label{s:intro}

There are many similarities between electromagnetic (\EM) radiation and
gravitational radiation: both travel at the speed of light; both carry
energy away from their sources; both consist of transverse waves with
two polarizations. In addition, Einstein's general relativity, the theoretical
underpinning of gravitational waves, can be put into a form remarkably 
parallel to Maxwell's electrodynamics, the theoretical underpinning of 
\EM\ waves. Despite the many similarities, there are important 
differences, and focusing on those differences helps to give a deeper understanding
of both kinds of radiation. 

Our mechanism for exploring those differences will be the visualization
of the fields.  Visualization of \EM\ field lines has proven very helpful
to student understanding \cite{johnsvispaper}. The
visualization of gravitational fields has been a challenge, but the
recently developed technique of using ``tendex'' and ``vortex'' 
lines\cite{CorntechPRL,CorntechPRD,CorntechZeroes}
provides insights that may be of pedagogical value comparable to the
plotting of electric or magnetic field lines in \EM.

Both in \EM\ and in gravitation, the dynamic nature of radiation
fields is of central importance to visualization, and in both \EM\ and
gravitation, waves carries energy. In \EM\ it will turn out that a
definite meaning can be given to the motion of field lines and to the
transport of energy. By contrast, in the gravitational case, a
definite meaning cannot be given to either of these concepts. This
will help us understand some important ways in which gravitation
fundamentally differs from E\&M.

The rest of this paper is organized as follows. In
Sec.~\ref{subsec:theoryem} we give a very brief review of \EM\ theory
to serve as a basis for comparison with the elements of gravitational
theory that we subsequently present in Sec.~\ref{subsec:theorygrav}.
Section \ref{s:statvis} develops the principles of visualization of
static fields, for E\&M in Sec.~\ref{sub:EM}, and for gravitaion in
Sec.~\ref{sub:GW}. In Sec.~\ref{s:statvis}, to illustrate both E\&M
and gravitational static fields, we focus on a particular model that
will be useful later, in the discussion of radiation. A dipole is the
simplest model for a source of E\&M radiation, but a gravitational
dipole cannot generate radiation. We therefore choose the simplest
configuraton that can generate both E\&M and gravitational waves: a
point quadupole. 

For simplicity of visualization, we eliminate the
issues of visualizing truly three dimensional fields by having the
point quadrupole be axisymmetric. Time changing sources give rise to 
radiation fields for both E\&M and gravitation. These dynamical 
fields and their visualization are discussed  in Sec.~\ref{s:dynavis}
with the example of oscillating E\&M and gravitational quadrupoles.
 We summarize and restate
our conclusions in Sec.~\ref{s:conc}.

A few words are in order about the choices that have been made for
notation and conventions.  The principle has been to produce a paper
that can be understood by a reader with a minimum of mathematical
preliminaries. To that end we have avoided certain practices that are
common in advanced literature. The following points deserve particular
notice. (i)~Papers involving relativity typically assume units in which the
$c$, the speed of light, is taken to be unity. In
order to have expressions in which the dimensionality of quantities is
more transparent, we do not make that choice; all factors of $c$
explicitly appear. (ii)~It is common to use the ``Einstein summation
convention,'' in which summation is assumed for any repeated
index. Adopting this convention would allow us to drop the explicit
summation symbols in Eq.~(\ref{Lorentzforce}) and many subsequent
equations. We have chosen, however, to have these summation symbols
appear. (iii)~To avoid the mathematical baggage of covariant
differentiation we have been explicit in using only Cartesian
coordinates and Cartesian components where expressions involve
differentiation, as in Eq.~(\ref{eq:EMcartcomps}). (iv)~We avoid
``coordinate bases'' that are commonly used in computations with
tensor fields. Rather, the components expressed, e.g., in
Eq.~(\ref{staticcalE}), are with respect to the familiar spherical
coordinate orthonormal basis, not the coordinate basis. One of the
simplifications following from this choice is that indices on components are the
same whether they are superscripts or subscripts; their location is
chosen for convenience.

\section{Electromagnetic and gravitational fields: introductory theory}
\label{s:theory}

\subsection{Electromagnetic fields}\label{subsec:theoryem}
For comparison with the gravitational case, it is useful for us to
mention the very roots of electromagnetic physics. The electric field
${\bf E}$ and magnetic field ${\bf B}$ are defined through the
expression for the acceleration {\bf a} due to the total electromagnetic
force, the Lorentz force, acting on a point particle of mass $m$ and
charge $q$, moving at velocity {\bf v},\cite{noSRT}
\begin{equation}\label{eq:Loraccel}
  {\bf a}=\frac{q}{m}\left({\bf E} +{\bf v}{\bs\times}{\bf B}\right)\,.
  \end{equation}
The fields  obey the Maxwell equations, which
in rationalized MKS units,  take the form
\begin{equation}
  \nabla\cdot {\bf E}=0\quad\quad
\nabla\times{\bf E}+\frac{\partial {\bf B}}   {\partial t}=0
\end{equation}
\begin{equation}
  \nabla\times{\bf B}-\frac{1}{c^2}\frac{\partial {\bf E}}{\partial t}=0
\quad\quad \nabla\cdot{\bf B}=0\ .
\end{equation}
Here we have simplified the equations by assuming that they apply in a
region devoid of sources and of material properties, i.e., we take
the charge and current density to be zero, and we assume vacuum values
of the dielectric constant and magnetic permeability.
It will be useful to rewrite these equations in component form (in 
which we assume a Cartesian basis):
\begin{equation}\label{Lorentzforce}
  a^j=\frac{q}{m}\left(E^j+\sum_{k,p}\epsilon_{jkp}v^kB^p\right)
\end{equation}

\begin{equation}\label{eq:EMcartcomps}
  \sum_{j}\frac{\partial E_j}{\partial x^j}=0\quad\quad
\sum_{j,k}\epsilon_{ijk}\,\frac{\partial E_k}{\partial x^j}
+\frac{\partial B_i}{\partial t}
=0
\end{equation}
\begin{equation}
  \sum_{j,k}\epsilon_{ijk}\,\frac{\partial B_k}{\partial x^j}-\frac{1}{c^2}\,
\frac{\partial E_i}{\partial t}
=0
\quad\quad   \sum_{j}\frac{\partial B_j}{\partial x^j}=0\,.
\end{equation}
Here the summations are over the indices of the Cartesian components
and coordinates, $\{x^1,x^2,x^3\}$, or $\{x,y,z\}$. The symbol $\epsilon_{ijk}$
is the three-dimensional alternating symbol used in the construction of 
determinants and cross products. It vanishes if any of its indices is 
repeated (e.g., $\epsilon_{221}=0$), equals +1 for any even permutation 
of 1,2,3, or $x,y,z$, (i.e., $\epsilon_{123}=\epsilon_{312}=\epsilon_{231}
=1
$,), and equals -1 for any odd permutation (i.e., $\epsilon_{213}=\epsilon_{321}=
\epsilon_{132}
=1
$).

The simplest solutions of the Maxwell equations are the time independent solutions,
especially the point multipole (point charge, dipole, quadrupole ...) solutions for electrostatic
or magnetostatic fields. Of greatest interest in this paper will be the not-so-simple
radiation solutions in which time variation is essential.
Of considerable importance to these radiation
solutions 
 is the Poynting vector, 
the flux of electromagnetic power per unit cross sectional area
\begin{equation}
  {\bf P}=\frac{1}{\mu_{0}}{\bf E}\times{\bf B}\,.
\end{equation}

In a  general treatment, the computation of the radiation produced by a given
distribution of time-changing charges and currents leads to retarded
integrals over those sources. Here, however,  we are primarily interested in the
description and visualization of the resulting fields, so we simply
invoke radiation fields without being specific about the internal
details of their source.  For such purposes the choice usually made is
that of a point dipole, but for our purpose here this choice
has the disadvantage that it has no gravitational analog. The lowest
order multipole for gravitational radiation is the
quadrupole\cite{whynodipole}. Accordingly, as our example 
of an electromagnetic radiating source we choose a point quadrupole, and we make
the description 
 and visualization as
simple as possible by taking the quadrupole to be axisymmetric.

\subsection{Gravitational fields}\label{subsec:theorygrav}
To understand what is meant by  ``gravitation'' in relativistic theories
it is best to start with the simplest case, gravitostatics, gravitation for
static configurations.
In this case, if the 
gravitational fields are typically weak (if they do not drive particles
to speeds comparable to $c$) then Newtonian ideas can be adopted, with minor 
modification, to relativistic gravitation.
By ``gravitational field,'' in this approach,  we do not mean, e.g., the downward
acceleration at 9.8\,m/sec$^2$ near the surface of the Earth. 
More generally, if 
 $\Phi_g$ is the usual Newtonian potential, 
$\nabla\Phi_g$ is not considered ``true'' gravitation. Since it affects
all particles identically, its effects on particles undergoing
their natural, freely falling motion disappear in a freely falling
frame, the inertial frame in the relativistic view of spacetime.

Gravitation, in the relativistic viewpoint, is the way in which the
the natural free-fall motions vary from place to place and time to
time.  In a static configuration, one in which there is no change from
time to time, the information is contained in the way in which
$\nabla\Phi_g$ varies from place to place. The rate of variation of a
vector is a tensor, in the case of $\nabla\Phi_g$ it is often called
the gravitoelectric field\cite{paradigm}.  (Because this tensor
describes the raising of tides on astrophysical objects in a Newtonian
setting it is also called the tidal tensor.)  In a Cartesian basis the
content of this tensor is the set of tensor
components\cite{relversion}
\begin{equation}\label{eq:gravitoelectricdef}
  {\cal  E}_{jk} =\frac{\partial^2\Phi_g}{\partial x^j\partial x^k}\,.
\end{equation}
The trace of this gravitoelectric tensor is a familiar quantity
\begin{equation}\label{eq:traceEkk}
  \sum_k {\cal  E}_{kk}=  \sum_k \frac{\partial^2\Phi_g}{\partial x^k\partial x^k}
=\nabla^2\Phi_g\,.
\end{equation}
Just as the electric field is divergenceless outside sources, the gravitoelectric
field is traceless outside sources. 

The definition in Eq.~(\ref{eq:traceEkk}) is valid even in the presence 
of sources. In that case the right hand side has the familiar value $4\pi G$
times mass density. This equation, then, 
is analogous to
Coulomb's law for electromagnetism. It gives a definition of the field
in terms of its sources, but only for a static field. To deal with
radiation we need more general definitions, definitions based on the
manifestations, the physical effects, of these fields. In the case
of electromagnetism, this more general definition is given by
Eq.~(\ref{eq:Loraccel}) or, equivalently,
Eq.~(\ref{Lorentzforce}).
In relativistic gravitation there is no 
concept of force {\em per se}. Rather, the manifestations of gravity
are seen in  the effects on  two point particles
separated by a small displacement ${\bf s}$, and with a relative
velocity ${\bf v}$. The relative acceleration of the two particles, given by\cite{geodev}
\begin{equation}\label{eq:geodev}
  \frac{d^2s^j}{dt^2}=-\sum_{k}{\cal E}_{jk}s^k-2\sum_{k,p,m}\epsilon_{jkp}{\cal
    B}_{pm}v^ks^m\,,
\end{equation}
defines the tensors $\calEbf$ and $\calBbf$.
The similarity to the Lorentz acceleration in
Eq.~(\ref{Lorentzforce}) is striking,
especially if one considers ${\cal E}_{jk}s^k$ and ${\cal B}_{jk}s^k$
to be vectors. There are, of course, differences of detail, one of
which is very fundamental. In Eqs.~(\ref{eq:Loraccel}), or (\ref{Lorentzforce}),
the factor $q/m$ describes the special features of the particle undergoing
electromagnetic acceleration. 
By contrast, in
Eq.~(\ref{eq:geodev}) there is no reference to any characteristic of
the particles.  In accordance with the so-called ``equivalence principle,''
gravitation acts in the same way on any particle.

The gravitoelectric and gravitomagnetic fields defined by
Eq.~(\ref{eq:geodev}) are symmetric ( ${\cal E}_{jk}={\cal E}_{kj}$,
${\cal B}_{jk}={\cal B}_{kj}$). Outside sources these fields are traceless
($\sum_k{\cal E}_{kk}$=$\sum_k{\cal B}_{kk}$=0 ) and obey the
Maxwell-like relations\cite{bianchiident}
\begin{equation}
  \sum_{j}\frac{\partial {\cal E}_{jk}}{\partial x^j}=0\quad\quad
\frac{1}{2}
\left(\sum_{jk}\epsilon_{pjk}\frac{\partial{\cal E}_{qk}}{\partial x^j}
+\sum_{jk}\epsilon_{qjk}\frac{\partial{\cal E}_{pk}}{\partial x^j}\right)
+\frac{\partial {\cal B}_{pq}}{\partial t}
=0
\end{equation}
\begin{equation}
\frac{1}{2}
\left(\sum_{jk}\epsilon_{pjk}\frac{\partial{\cal B}_{qk}}{\partial x^j}
+\sum_{jk}\epsilon_{qjk}\frac{\partial{\cal B}_{pk}}{\partial x^j}\right)
-\frac{1}{c^2}\,\frac{\partial {\cal E}_{pq}}{\partial t}
=0
\quad\quad   \sum_{j}\frac{\partial{\cal B}_{jk}}{\partial x^j}=0\,,
\end{equation}
which can 
be written as
\begin{eqnarray}
 {\bs \nabla\cdot{\bs{\cal E}}}=0\quad\quad\quad
{\bs\nabla\times{\bs{\cal E}}}+\frac{\partial{\bs{\cal B}}}{\partial t}=0\\
{\bs\nabla\times{\bs{\cal B}}}-\frac{1}{c^2}\,\frac{\partial{\bs{\cal E}}}{\partial t}=0
\quad\quad\quad {\bs \nabla\cdot{\bs{\cal B}} }=0\,,\label{eq:curlyBeqs}
\end{eqnarray}
with appropriate interpretation of the divergence and curl. Note that 
the divergence can be taken on either index (since the ``gravito-'' tensors
are symmetric) and the curl in these equations is symmetrized\cite{c2factor}.

Two ``theoretical'' points bear mentioning: (i) Just as the six independent 
components of ${\bf E}$ and ${\bf B}$ contain a complete description of the 
electromagnetic field at a point, the ten independent components of the
two symmetric traceless tensors ${\bs{\cal E}}$ and ${\bs{\cal B}}$
contain a complete description of the gravitational
field at a point. (ii) The vectors ${\bf E}$ and ${\bf B}$ are ``gauge invariant.''
They cannot, for instance, be made to vanish by a mathematical choice. In the 
same sense ${\bs{\cal E}}$ and ${\bs{\cal B}}$ are gauge invariant\cite{vsmetricperts}.
This is the mathematical equivalent of the physical statement that these quantities 
are directly physically measurable.


\section{Visualization of static electromagnetic and gravitational fields}
\label{s:statvis}

\subsection{Electromagnetism}\label{sub:EM}

\subsubsection{General considerations for electromagnetic visualization}\label{subsec:genconsidEM}
The electric and magnetic fields are vector fields, and hence in
principle are simple to picture. One can use small arrows indicating
field direction, with arrow length indicating vector
magnitude.  Alternatively one can sketch field
lines, curves to which the vectors are tangent, with the density of
these lines indicating the strength of the field. The limitations of
spatial
resolution in these methods are well known. Below we will be
illustrating vector and line fields with a more modern technique: the line
integral convolution (LIC) method of Cabral and Leedom\cite{LIC}.  In
this method the brightness or darkness of pixels is correlated along field lines.
The method produces images with streaks showing the structure of the
field lines in an intuitively appealing way and with resolution approaching 
that of the display.

\subsubsection{Static electromagnetic point quadrupole }

\begin{figure}[htb]
\includegraphics[height=1.5in]{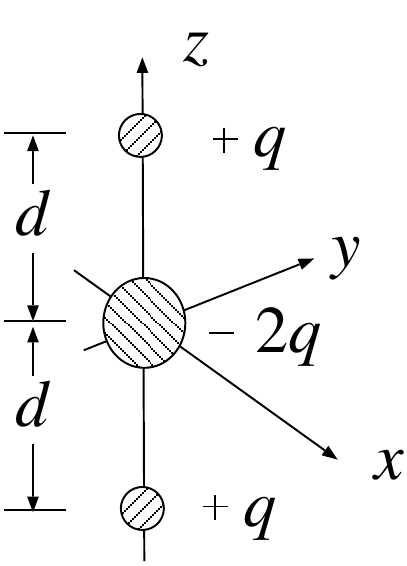} \caption{
A simple model of an electric quadrupole.
\label{fig:elecstatquad}}
\end{figure}

As the simplest example of an 
electromagnetic multipole source that will be generalizable to gravitation, 
we choose an axisymmetric quadrupole.
A  realization of such a source is shown in
Fig.~\ref{fig:elecstatquad}: two equal positive charges symmetrically
arranged on the $z$ axis about a double negative charge at the
origin. Since there is no net charge, and no favored positive
direction, the configuration has neither a monopole nor a dipole
moment. The static configuration pictured then has a quadrupole
as its lowest nonvanishing multipole.

In general, the Cartesian components of an electric quadrupole are
given 
by\cite{JacksonQ}
\begin{equation}\label{quadelecdef}
  Q_{ij}=\int (3x_ix_j-r^2\delta_{ij})\rho({\bf x}) d^3x\,,
\end{equation}
where $\rho({\bf x})$ is charge density.
For our model in
Fig.~\ref{fig:elecstatquad}, the specific components are
\begin{equation}\label{quadeleccomps}
  Q_{zz}=4d^2q\equiv4Q\quad\quad Q_{xx}=  Q_{yy}=      -2Q\,,
\end{equation}
with $Q_{ij}=0$ for $i\neq j$.
Note that the quadrupole $Q_{ij}$ is itself a tensor describing 
the charge distribution of the source. It is not a tensor field.
The electric {\em field} produced by  that source is the vector field {\bf E}.

The model in Fig.~\ref{fig:elecstatquad} is
not a ``pure'' quadrupole; it has multipole moments of order 2, 4,
6\ldots. More important, it is not a ``point source''; it has a
characteristic size $d$.  To create a pure point quadrupole we use a
limiting process analogous to that for defining a point dipole.  We
shrink $d$ to zero, while keeping finite the product $Q\equiv d^2
q$. The result is a point source with only a quadrupole moment.

\begin{figure}[htb]
\includegraphics[width=.35\textwidth]{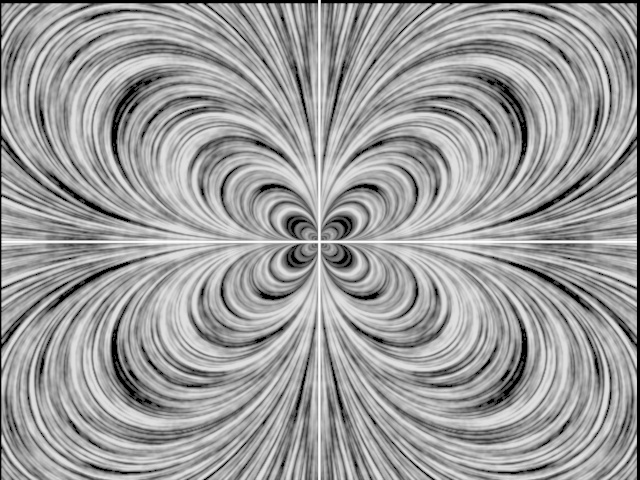}\hspace{12pt} 
\caption{A LIC of the electric field lines of the 
azimuthally symmetric static electric quadrupole
described in the text.\label{fig:LICestat}}
\end{figure}

The static electric field for the point quadrupole is most simply
computed from the electrostatic potential. A formal
procedure\cite{JacksonQ} can be used to find the potential directly
from the quadrupole components in Eq.~(\ref{quadeleccomps}), or a
limiting procedure can be applied to the potential of the three point
charges in Fig.~\ref{fig:elecstatquad}. The resulting electrostatic potential 
is
\begin{equation}
  \Phi_e=\frac{Q}{4\pi\epsilon_0 r^3}\left(3\cos^2\theta-1\right)\,,
\end{equation}
and hence the electric field ${\bf E}=-{\bs\nabla}\Phi_e$
has spherical components
\begin{equation}\label{statelec}
  E_r=\frac{6Q}{4\pi\epsilon_0r^4}\left(\frac{3}{2}\cos^2\theta-\frac{1}{2}\right)
\quad\quad\quad\quad E_\theta=\frac{6Q}{4\pi\epsilon_0r^4}
\cos\theta\,\sin\theta\ .
\end{equation}
The electric field topology for this case is illustrated in
Fig.~\ref{fig:elecstatquad}. For this static electric configuration,
there are no associated magnetic fields.

\subsection{Gravitation}\label{sub:GW}
\subsubsection{General considerations for gravitational visualization}

The problem of visualizing the  ${\cal E}_{jk}$
and ${\cal B}_{jk}$ fields is a special example 
of the question: how does one visualize tensorial 
fields.  
The ``gravito-'' fields, ${\cal E}_{jk}$ and ${\cal B}_{jk}$, are
tensors, but at least they are the simplest nontrivial type of
3-dimensional tensor: they are second rank (two index) symmetric
tensors. In this sense they are similar to the most familiar tensors
of physics: the inertia tensor, the stress tensor, the dielectric
tensor of an anisotropic material, etc.

In the case of a tensor like the inertia tensor of an extended massive
object, a reasonable visualization is the inertia ellipsoid, a three
dimensional ellipsoid whose shape shows the directions and size of the
principal axes of the moment of inertia of the massive
object\cite{allpositive}. This ellipsoid for a second-rank symmetric 
tensor is very much the analog of the arrow for a vector.

The moment of inertia ellipsoid describes a single tensor, not a
tensor field. While a display of space filled with arrows has some
usefulness for the visualization of a vector field, the same is probably
not true of space filled with ellipsoids.  But, just as arrows can be
connected together to form field lines, the {\em principal axes} of
tensorial ellipsoids can be connected to form a network of lines with
visualization properties somewhat similar to field lines.  The
specific technique for visualization of a symmetric tensor $A_{jk}$ is
to find, at each point in space, the principal directions, i.e., the
vectors $v^k$ satisfying the eigenvector condition
$A_{jk}v^k=\lambda \,v^k$. If $A_{jk}$ is symmetric we are
guaranteed that three such eigenvectors exist and are orthogonal. The three
orthogonal sets of lines connecting these eigenvectors, the
``eigenlines'' for these eigenvectors then give us  visual
information about  our tensor field.

This method has been used for some time in the 
visualization of stress (a tensor quantity) in 
fluid
dynamics and solid mechanics\cite{stress}, and has
recently been suggested for use in visualizing the gravitoelectric and
gravitomagnetic fields\cite{CorntechPRL,CorntechPRD,CorntechZeroes}.
Those advocating the application to gravitation give the name ``vortex'' lines to the 
eigenvectors of ${\bs{\cal B}}$ due to the role of that field in driving 
the precession of spin. The 
eigenvector field lines of 
${\bs{\cal E}}$ are given the name ``tendex'' lines, suggestive of
the role of the tidal distortions associated with ${\bs{\cal E}}$.
This approach to visualization 
holds the promise of giving the kind of intuitive insights that may 
suggest what configurations lead to strong emission of power and of
linear momentum in gravitational waves. 

\subsubsection{Static gravitational point quadrupole}
We now consider the gravitational equivalent of the configuration in
Fig.~\ref{fig:elecstatquad}. Here the objects at $z=\pm d$ are points of
mass $M$, rather than points of charge $q$. In analogy
with  Fig.~\ref{fig:elecstatquad}, we include
a negative mass $-2M$ at the center\cite{neg2M}.  
\begin{figure}[htb]
\includegraphics[width=.15\textwidth]{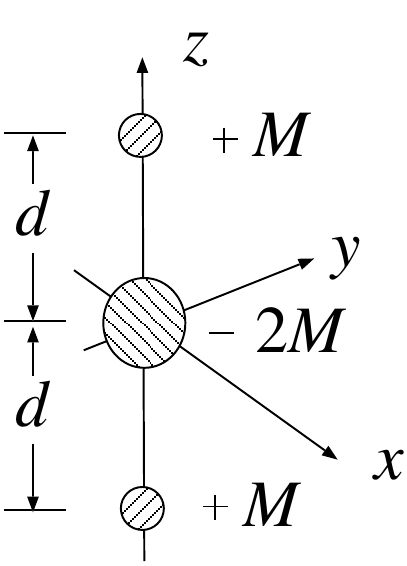} \caption{
A simple model of a gravitational quadrupole.
\label{fig:gravquad}}
\end{figure}
The quadrupole components for the gravitational case are given by the
same integral as that in Eq.~(\ref{quadelecdef}), but with $\rho$ now
representing mass density. The components therefore are those in
Eq.~(\ref{quadeleccomps}) with $q$ replaced by $M$.

The Newtonian gravitational potential for the mass configuration in
Fig.~\ref{fig:gravquad} is the same as the
electrostatic potential of the charge configuration in Fig.~\ref{fig:elecstatquad}
after the replacements $q\rightarrow M$ and $1/4\pi\epsilon_0\rightarrow -G$.
The limit of this potential, for $d\rightarrow0$ with $Q=Md^2$ fixed, is
\begin{equation}
  \Phi_g=-\frac{GQ}{r^3}\left(3\cos^2\theta-1\right)\,.
\end{equation}

The components of the static gravitoelectric field are straightforward
to compute from this potential\cite{fieldcalc}.  For the static
gravitational quadrupole, in analogy with the electric case, there are
no gravitomagnetic fields\cite{whynocalB}. The components of the
gravitoelectric fields, in a spherical basis, are given by
\begin{eqnarray}\label{staticcalE}
  \calerr&=&
-\frac{12GQ}{r^5}\left(3\cos^2\theta-1\right)\nonumber
\\
  \calert&=&-\frac{24GQ}{r^5}\sin\theta\cos\theta\label{eq:calEstatcomps}      \\
\calepp-\calett&=&\frac{6GQ}{r^5}\,\sin^2\theta\nonumber
\,.
\end{eqnarray}
The components ${\cal E}_{r\phi}$ and ${\cal E}_{\theta\phi}$ are zero
by axisymmetry, and the individual components $\calepp,\calett$ follow
from the last of Eqs.~(\ref{staticcalE}) and from the tracelessness
condition $\calepp+\calett=-\calerr$.

\begin{figure}[htb]
\hspace{0in}\includegraphics[width=.4\textwidth]{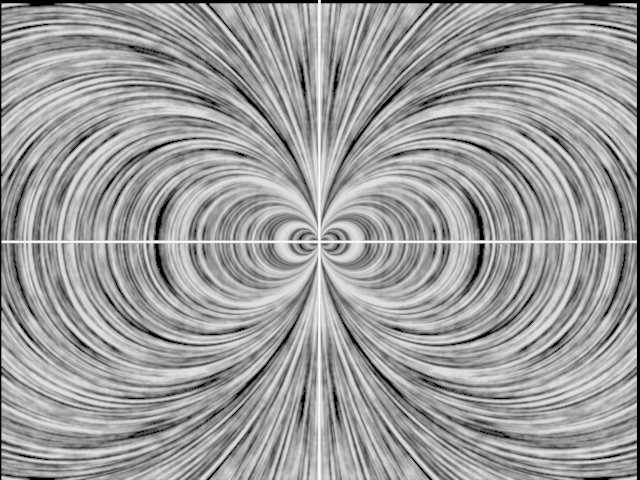}\hspace{12pt}\\

\vspace{.03in}
\hspace{0in}\includegraphics[width=.4\textwidth]{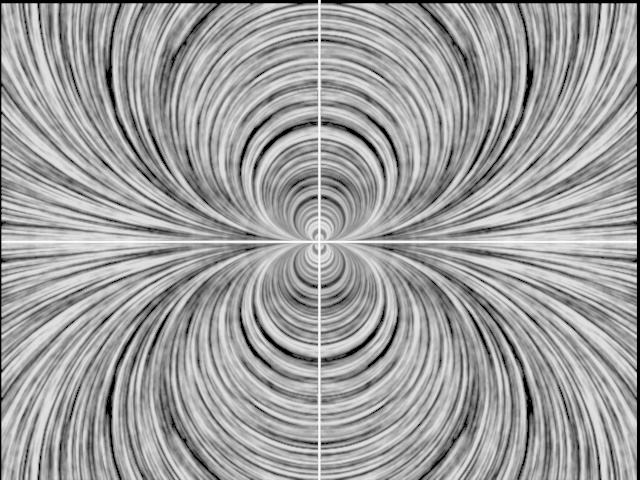}\\ \vspace{.5in}
\caption{ Line Integral Convolutions showing the eigenvector fields
  for the vertically oriented static gravitational point quadrupole.
  The top image shows the field for negative eigenvalues, and the
  bottom image shows the field for positive eigenvalues.
\label{fig:statgraveigens}}
\end{figure}

We now turn to the issue of visualizing these fields. The structure of 
the components of $\calEbf$ shows that two of the eigenvectors will be in 
the $r,\theta$ plane, and one in the $\phi$ direction.
The eigenlines in the $\phi$ direction are simple azimuthal circles. The 
much more interesting eigenlines in the 
$r\theta$ plane are  shown, as LIC images, in Fig.~\ref{fig:statgraveigens}.

\vspace{.2in}

The comparison of Fig.~\ref{fig:statgraveigens} and
Fig.~\ref{fig:LICestat} is very instructive
and poses a sequence of questions.
The electrostatic and
gravitostatic potentials are identical aside from trivial
replacements. Why are the visualizations so different?
The immediate answer is that in the electromagnetic 
case we are picturing 
${\bf  E}=-{\bs\nabla}\Phi_e$, a vectorial quantity with components
$\partial\Phi_e/\partial x^j$. In the gravitational case we are 
picturing not the vector field ${\bf  g}=-{\bs\nabla}\Phi_g$, with components
$\partial\Phi/\partial x^j$, but 
the tensorial quantity ${\bs{\cal E}}$ with the components.
$\partial^2\Phi_g/\partial x^j\partial x^k$. The component notation 
correctly suggests that  ${\bs{\cal E}}$ is the  gradient\cite{neggrad} of 
${\bf  g}$. Why not, then, simply visualize gravitational fields
with images of ${\bf  g}$? For that matter why not simply use
potentials in both cases? 

In electromagnetism we know the answer. The electrostatic potential is
useful for electrostatics, but the concept does not carry over to
time-changing fields, and hence to radiation. For electromagnetic
radiation the electric field ${\bf E}$ is a valid and important
concept, but it is not simply the gradient of a scalar.  The same
turns out to be true in gravitation.  For gravitational radiation
${\bs{\cal E}}$ is a valid and important concept, but its components
are not the second derivatives of a scalar field.

\section{dynamic electromagnetic and gravitational fields and their visualization}
\label{s:dynavis}

\subsection{Electromagnetic fields}
\label{sub:visdynaEM}
\subsubsection{General visualization considerations}

While LIC snapshots are very useful, they show the fields only at a
moment of time, and do not capture the necessarily dynamic nature of
the radiation fields. To address this, a new technique, Dynamic Line
Integral Convolution  (DLIC), has been
developed\cite{johnsvispaper,Sundquist}.  Quite aside from the technical
challenges to be overcome is a fundamental question underlying the
notion of field line movement. Animations can show how a certain
field line changes from one moment to the next, but what do we mean by
``a certain field line''? How do we relate the field line at one
moment to the ``same'' field line at another moment? How do we put
unchanging ``tags'' on a field line\cite{BnO}. It turns out that there
is a reasonable criterion for tagging field lines. At least in the
case of axisymmetric sources, the mathematical nature of Maxwell's
equations imposes these tags in a natural way. Furthermore, the same tagging
can be justified on purely physical grounds.

\begin{figure}[htb]
\includegraphics[width=.35\textwidth]{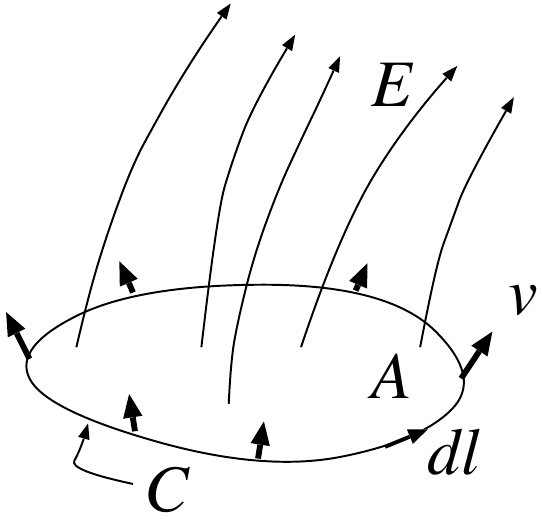} \caption{
Electric flux ${\bf E}$ through an area whose boundary is moving
with velocity ${\bf v}$.
\label{fig:Eflux}}
\end{figure}
The mathematical property is illustrated in Fig.~\ref{fig:Eflux}. The
flux of electric field through an area $A$, with unit normal ${\bf n}$, 
changes in time according to
$$
  \frac{d}{dt}\int_{A} {\bf E}\cdot{\bf n}\,dA=\int_A 
\frac{\partial{\bf E}}{\partial t}\cdot{\bf n}\,dA+\int_C \left({\bf E}\times{\bf v}\right)\cdot d{\bf l}
=\int_A 
c^2\nabla\times{\bf B}\cdot{\bf n}\,dA
+\int_C \left({\bf E}\times{\bf v}\right)\cdot d{\bf l}
$$\begin{equation}
=\int_C \left(c^2{\bf B}+{\bf E}\times{\bf v}\right)\cdot d{\bf l}\,.
\end{equation}
We adopt flux conservation as our criterion for the motion of electric field lines. 
This requires that the field line velocity ${\bf v}$
satisfy
\begin{equation}\label{Evconstraint}
  c^2{\bf B}+{\bf E}\times{\bf v}=0\,.
\end{equation}
For magnetic field lines, the equivalent condition is 
\begin{equation}\label{Bvconstraint}
  {\bf E}+{\bf v}\times{\bf B}=0\,.
\end{equation}
The component of ${\bf v}$ along the field lines is meaningless, so
these conditions give one constraint on the components of ${\bf v}$
perpendicular to the field lines. In the case of axisymmetry, to be
considered below, that is all we need.

For a magnetic field configuration, e.g., an oscillating magnetic
dipole, we can interpret Eq.~(\ref{Bvconstraint}) as the condition
that a charged particle experience no net Lorentz force.  Particles
trapped in tight orbits around magnetic field lines {\em do} in fact,
move with field lines, so this condition has a very practical meaning
in many plasma situations: the motion of field lines is equivalent to
the motion of electrons trapped on tight orbits around field lines.
For an electric field these same considerations apply if we replace
the electric monopole charge motion with the motion of a
(hypothetical) magnetic monopole charge.

In the radiation field, and in many nonradiative configurations, the electric and 
magnetic fields are orthogonal, i.e., ${\bf E\cdot B}=0$. In this case the solution 
to Eqs.~(\ref{Evconstraint}) and (\ref{Bvconstraint}) are, respectively,
\begin{equation}\label{eq:vandPoynting}
  {\bf v}=c^2\frac{{\bf E\times B}}{E^2}
\quad\quad\quad
  {\bf v}=\frac{{\bf E\times B}}{B^2}\,.
\end{equation}
It is interesting that the numerator in both cases is proportional
to the Poynting flux, indicating that the flux-conserving motion of the field 
lines is compatible with the transport of electromagnetic energy.

\subsubsection{Oscillating electric point quadrupole}

We can convert the static quadrupole of Fig.~\ref{fig:elecstatquad} to a quadrupole
source oscillating at frequency $\omega$
by replacing  the static distances $d$ by the oscillating
distances $d_0+\Delta d \cos{\omega t}$. The ``amplitude'' of 
the quadrupole then changes from $Q=qd^2$ to 
\begin{equation}
  Q(t)=q(d_0+\Delta d\,\cos{\omega t})^2=qd_0^2+2qd_0\Delta d\cos{\omega t}
+q(\Delta d\cos{\omega t})^2\,.
\end{equation}
We now take only the part of this expression that oscillates at
frequency $\omega$, and we define $Q\equiv2qd_0\Delta d$.  With this
meaning for $Q$, the spherical components of fields produced by
this source are\cite{EMcalcs}
\begin{equation}\label{eq:Erdynamic}
E_r=-2\frac{Q k^2}{4\pi\epsilon_0\,r^2}\left[
\cos{(kr-\omega t)}\left(1-\frac{3}{k^2r^2}\right)
-\frac{3}{kr}\sin{(kr-\omega t)}
\right]
\left(\frac{3}{2}\cos^2\theta-\frac{1}{2}\right)  
\end{equation}
\begin{equation}\label{eq:Ethetadynamic}
E_\theta=-\frac{Q k^2}{4\pi\epsilon_0\,r^2}\left[
\sin{(kr-\omega t)}\left(kr-\frac{6}{kr}\right)
+\left(3-\frac{6}{k^2r^2}\right)
\cos{(kr-\omega t)}
\right]
\cos\theta\,\sin\theta\,,  
\end{equation}
\begin{equation}\label{eq:Bphidynamic}
  B_\phi=-\frac{Q k^3}{4\pi\epsilon_0 c\,r}\left[
\left(1-\frac{1}{3k^2r^2}\right)\sin{(kr-\omega t)}+\frac{3}{kr}\cos{(kr-\omega t)}
\right]\cos\theta\,\sin\theta\,,
\end{equation}
where $k\equiv\omega/c$.

 In the limit $kr\rightarrow0$ the dominant terms in the field
are proportional to $1/r^4$ and these terms agree precisely with
those in Eqs.~(\ref{statelec}). This leads to an
 important insight about visualization: As $kr$ becomes much smaller
 than unity, the dominant terms in the field are the $1/r^4$ terms,
 and these terms have the form of the static solution modulated by
 $\cos{\omega t}$.  In other words, at distances from the source much
 less than a wavelength the field is a quasistatic field, a field with
 the structure of the static field, but with an amplitude oscillating
 in time.

Far from the source, in the region $kr\gg1$, the story is very different.
Here the radial electric field falls off as $~1/r^2$.
The   
 dominant fields are the $~1/r$ parts of $E_\theta$ and $B_\phi$:
\begin{equation}\label{eq:Erad}
  E_\theta=-\frac{Q k^3}{4\pi\epsilon_0\,r}
\sin{(kr-\omega t)}
\cos\theta\,\sin\theta
\quad\quad\quad
  B_\phi=-\frac{Q k^3}{4\pi\epsilon_0 c\,r}
\sin{(kr-\omega t)}
\cos\theta\,\sin\theta  \quad\quad \mbox{for $kr\gg1$}\,.
\end{equation}
These are the radiation fields, the fields in the limit that $r$ is much
larger than the radiation wavelength. The fact that 
the radial component does not contribute to this field is simply a
statement that the radiation fields are transverse, orthogonal to the 
direction to the source.

\begin{figure}[htb]
\includegraphics[width=.5\textwidth]{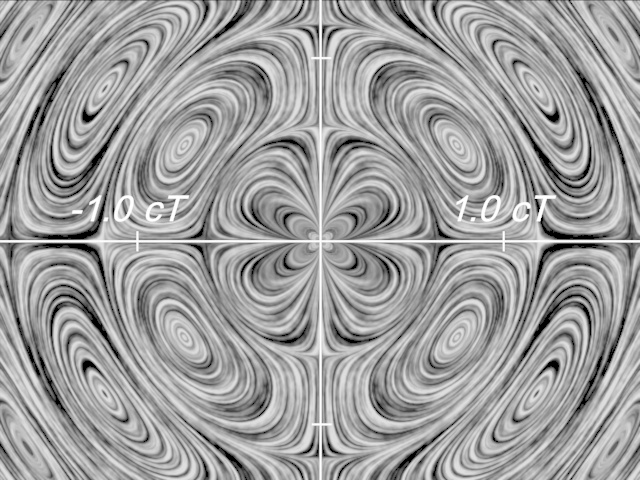} 
\caption{A Line Integral Convolution snapshot
of the electric field given, at time $t=0$ by Eqs.~(\ref{eq:Erdynamic}) and (\ref{eq:Ethetadynamic})
for an oscillating electric quadrupole.
The markers show points at one wavelength on either side of the origin.
At distances from the origin small compared to one wavelength $cT$ the solutions 
approach those of the static solutions in Fig.~\ref{fig:LICestat}.
This is one frame of a complete movie that can be found at
{\tt http://web.mit.edu/viz/gravrad/visualizations/EM/EMintermediate/}.
\label{fig:emradquad}}
\end{figure}

Figure~\ref{fig:emradquad} gives a LIC snapshot of the fields of the
  oscillating electric quadrupole at a single instant of time.  
The wavelength $\lambda=c/\omega=cT$ of the radiation is indicated on the figure, 
and it is particularly interesting to note that the central region of 
Fig.~\ref{fig:emradquad}, the region small compared to $\lambda$, is 
indistinguishable from 
Fig.~\ref{fig:LICestat}. 
In the opposite limit, 
the regions far from the central region, to the extent
that these are included in 
Fig.~\ref{fig:emradquad}, show the tendency to take the radiation form, with transverse 
field lines. 
What is most interesting in 
Fig.~\ref{fig:emradquad}, however, is neither the near or the far regions, but rather 
the intermediate region.
The figure reveals details of the structure of the transition fields
  that cannot easily be inferred from the mathematical expressions in
  Eqs.~(\ref{eq:Erdynamic}) and (\ref{eq:Ethetadynamic}). It shows how the field
structure makes the transition from the very different nature of the near and 
the far fields\cite{johnsvispaper}.

A snapshot of a dynamic field, however, necessarily shows only a
single phase of the radiation field. A full description requires an
animation, and an animation requires a way of identifying field lines
at different moments of time, as described in Sec.~\ref{sub:EM}. The
identification principle introduced in that subsection gives a single
constraint on the velocity of field lines orthonognal to the line
itself. In the axisymmetric case, geometry gives us the other
constraint. In particular, the electric field lines in
Fig.~\ref{fig:elecstatquad}, by symmetry, can have no $\phi$ component. The
condition in Eq.~(\ref{Evconstraint}) then completely fixes the line
velocity ${\bf v}$. A DLIC animation of the dynamic electric 
field lines is available online\cite{onlineDLIC}.

\subsection{Gravitational  fields}
\label{sub:visdynaGrav}

\subsubsection{Oscillating gravitational point quadrupole}

To create a point quadrupole gravitational source oscillating at
frequency $\omega$, we can perform precisely the same procedure as for
the electromagnetic case. We take the distances $d$ to have the form
$d_0+\Delta d \cos{\omega t}$ and we take the symbol $Q$ to mean
$2Md_0\Delta d$.
The nonvanishing spherical components of ${\bs{\cal E}}$
and ${\bs{\cal B}}$, 
the analogs of Eqs.~(\ref{eq:Erdynamic}), (\ref{eq:Ethetadynamic}), 
are found to be
\begin{eqnarray}
  \calerr&=&\frac{-4GQ k^2}{r^3}\,\left[\left(-1+\frac{3}{k^2r^2}\right)\cunew
+\frac{3\sunew}{ kr}\right]\left(3\cos^2\theta-1\right)\label{eq:calerr}\\
  \calert&=&\frac{-4GQ k^2}{r^3}\,\left[\left(\frac{6}{(kr)^2}-3\right)\cunew
-\left( kr-\frac{6}{ kr}\right)\sunew\right]\sin\theta\cos\theta\label{eq:calert}\\
  \calbrp&=&\frac{-4GQ k^2}{c r^3}\,\left[-{3}\cunew
-\left(k r-\frac{3}{kr}\right)\sunew\right]\sin\theta\cos\theta\label{eq:calbrp}\\
\calepp-\calett&=&\frac{-2GQ k^2}{r^3}\,\left[\left(
-\frac{3}{k^2r^2}+3-(k r)^2
\right)\cunew
-\left(
\frac{3}{kr}-2k r
\right)\sunew\right]\sin^2\theta\label{eq:calepp}\\
\calbtp&=&\frac{-GQ k^2}{c r^3}\,\left[\left(
-3+k^2r^2
\right)\cunew
-\left(
-\frac{3}{kr}+2kr
\right)\sunew\right]\sin^2\theta\label{eq:calbtp}\,.
\end{eqnarray}

As was the case for the dynamical fields in Eqs.~(\ref{eq:Erdynamic}),
(\ref{eq:Ethetadynamic}) there are interesting limits to these
expressions for both $kr\ll1$ and $kr\gg1$. In the former case, we find
that to the leading $1/r^5$ order
${\cal E}_{rr}$, ${\cal E}_{r\theta}$, and $\calepp-\calett$, have the
form of the values in Eqs.~(\ref{eq:calEstatcomps}), modulated by
$\cos{\omega t}$, while the gravitomagnetic components vanish to this order.
As is the case in electromagnetism, at distances from the source much less than 
a wavelength the fields are those of a quasistatic source, fields with the 
structure of a static field, but oscillating in time.

In the opposite limit,
for $kr\gg1$, the dominant components, those that fall off as $1/r$, are
\begin{eqnarray}
  \calepp=-\calett&=&\frac{GQ\omega^4}{r}\cunew\sin^2\theta\label{eq:epprad}\\
\calbtp&=&-\frac{GQ\omega^4}{r}\cunew\sin^2\theta\,.\label{eq:btprad}
\end{eqnarray}
The form of these radiation fields suggests transverse fields, a
suggestion that can be confirmed with the mathematics of general
relativity and an appropriate set of definitions and constraints. For
our purposes here it is sufficient to note that the eigenvector
directions can immediately be inferred.  From Eq.~(\ref{eq:epprad}),
and treating $\calerr$ as zero, we have that the matrix of
gravitoelectric field components is diagonal and hence that there are
eigenvectors of the gravitoelectric field in the $\theta$ direction, in
the $\phi$ direction and (for zero eigenvalue) in the $r$ direction.  From
Eq.~(\ref{eq:btprad}) we infer that in the radiation zone there are
two eigenvectors for the gravitomagnetic field in the $\theta\phi$
plane, each at 45 degrees from the $\theta$ and the $\phi$
directions. Again, there is an eigenvector field, for zero eigenvalue,
in the radial direction.

\begin{figure}[htb]
\includegraphics[width=.45\textwidth]{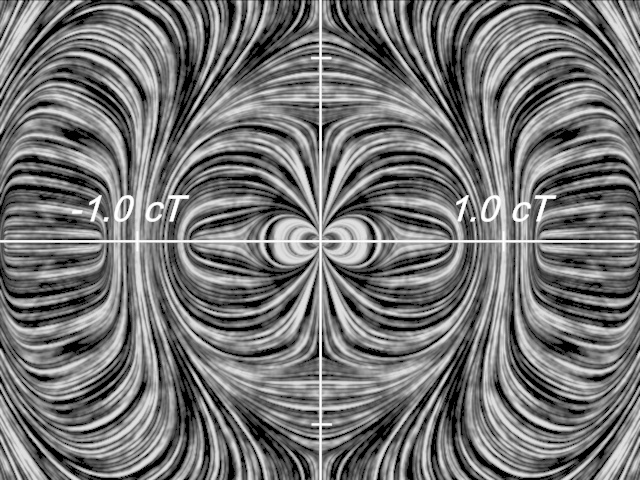}\\
\includegraphics[width=.45\textwidth]{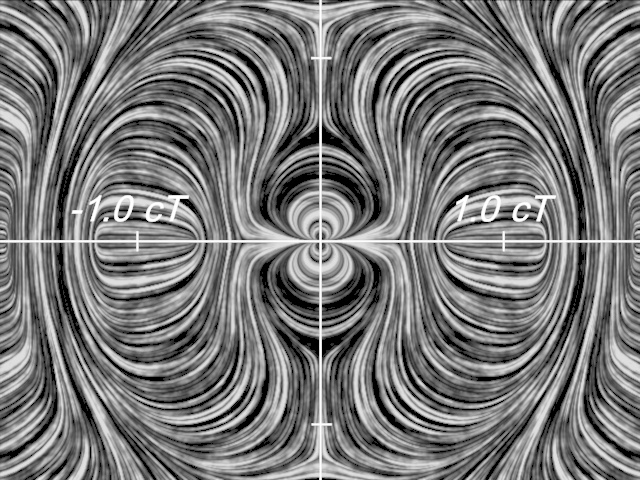}
 \caption{LIC snapshots of the eigenlines of the gravitoelectric field
${\bs{\cal E}}$
for an oscillating point 
gravitational quadrupole described by 
 Eqs.~(\ref{eq:calerr}), (\ref{eq:calert}) and
  (\ref{eq:calepp}) at time $t=0$.
The markers show the points at a distance of one wavelength on either side of
the origin.
%
The top LIC shows the eigenlines for one family\cite{family} of eigenvectors; the
LIC on the bottom shows the other family.  At distances from the
origin small compared to the wavelength $cT$, the fields, and hence
the eigenlines, approach those of the two families of eigenlines 
for the 
static solution shown in
Fig.~\ref{fig:statgraveigens}.
The top (bottom) image is one frame of a complete movie that can be found at
{\tt http://web.mit.edu/viz/gravrad/visualizations/GRAVinter/EposgravNoBkInt/}
({\tt http://web.mit.edu/viz/gravrad/visualizations/GRAVinter/EneggravNoBkInt/}
).
\label{fig:LICGsnapshot}}. 
\end{figure}

Snapshots of the eigenlines of ${\bs{\cal E}}$ are presented in
Fig.~\ref{fig:LICGsnapshot}. As in the electromagnetic case, these
snapshots show what we already learned from the mathematics. Close to
the source the eigenline field is quasistatic, and far from the source
it is transverse. Also as in the electromagnetic case, these figures
can clarify what the mathematics cannot: 
the form of the fields in the intermediate zone, at
distances from the source comparable to the wavelength $\lambda=
cT$. In this zone the fields must make a transition from the
quasistatic small-$r$ form to the radiating large-$r$ form.

Both in the mathematics and in Fig.~\ref{fig:LICGsnapshot} we see that
the form of the fields in the intermediate zone is even richer for the
oscillating gravitational quadrupole than for the oscillating electric
quadrupole.  In the latter case the radiation fields fall off as
$1/r$, while the near-source quasistatic field has the character
$1/r^4$. For the gravitational quadrupole the radiation fields have
the same $1/r$ character, but the near-source quasistatic fields
behave as $1/r^5$.

There is an interesting feature of the gravitational fields that is associated with
the mathematical expressions more than the visualizations. The
expressions in Eqs.~(\ref{eq:Erdynamic}), (\ref{eq:Ethetadynamic}),
and (\ref{eq:Bphidynamic}), for $E_r, E_\theta, B_\phi$, are identical
to those for $\calerr,\calert,\calbrp$ when the change
$1/4\pi\epsilon_0 r^2 \rightarrow -2GQ/r^3 $ is made; this is not true
only in the near-source zone or the radiation zone, but for all values
of $kr$! This means that, aside from a change in constants and a
single factor of $r$, the electromagnetic solution, including
radiation fields, is completely contained in the nonradiative part of
the gravitational solution. Stated in the other direction: the electromagnetic
solution  contains all of the gravitational solution except for those 
components 
$\calepp-\calett$, and $\calbtp$, that carry 
radiation (cf.~Eq.~(\ref{eq:epprad}) and
(\ref{eq:btprad})).

This, of course, is not a coincidence, nor is it an idiosyncrasy of
the axially symmetric point quadrupole. Rather it is a consequence of
the structure of classical theories that describe massless fields,
i.e., fields that propagate at the speed of light. This structure is
most apparent when the fields are described with the appropriate
mathematics, a set of scalars that result from projecting the fields
onto a set of basis vectors best suited to the analysis of propagating
fields\cite{NP}.  

This relationship of the electromagnetic and
linearized gravity fields gives us an additional possibility for
visualization. If we project the gravitational tensors with the unit
$r$ vector (equivalently take the dot product of $\calEbf$ and
$\calBbf$ with the unit radial vector $\bs{\hat{r}}$) we get two
vector fields, one with $r,\theta,\phi$ components
$\calerr,\calert,\calerp$, the other with components
$\calbrr,\calbrt,\calbrp$. The {\bf E} field differs from the vector
with components $\calerr,\calert,\calerp$ only by a factor of $r$, and
hence the direction of the two fields is the same at any point.  The
same is true for the corresponding {\bf B} field. The LICs of the {\bf
  E} and {\bf B}, as in e.g., Fig.~\ref{fig:emradquad}, fields can
therefore be considered to be LICs of all parts of the gravitational
field except the parts transverse to the $r$ direction. It is those
transverse parts, of course, that carry the radiation.

\subsubsection{General visualization considerations}

In linearized gravity, there is no equivalent of the principles that
define or constrain field line motion in electromagnetism. What this
means is that we can present a sequence of eigenlines, but we cannot
say which line at one moment corresponds to which line at
another. This fundamental problem turns out not to be a barrier if we
only want to show the qualitative nature of the field pattern motions
in gravitation.
 Intuitively useful 
flow fields can be guessed at qualitatively
due to the
existence of singularities.

In plots of lines tangent to vector fields, the singularities are
points at which the vector has zero magnitude, so that the direction
of a tangent line is undefined.  In Fig.~\ref{fig:emradquad}, such
points can be seen both in the equatorial plane and along the symmetry
axis. If the vector field illustrated were the velocity of a fluid,
these singular points would be called stagnation points.

It is simple to argue that lines of the electric field {\bf E} 
in electromagnetic quadrupole radiation
moving according to the
rules of Eq.~(\ref{Evconstraint}) do not cross
singularities. 
Similary, in the gravitational case,
singularities are easily identified, and the motion of
the singularities -- their displacement from one snapshot to another
-- therefore gives us a coarse visual sketch of the motion of the
field lines; once the position of the singularities is known at a new
time, the remainder of the field lines can be approximately drawn
in. Different choices of the details of how they are drawn make no
difference in the qualitative content of the images.

Note that the eigenline fields 
of gravitational radiation shown in  Fig.~\ref{fig:LICGsnapshot}
are          
richer in singularities than the electromagnetic vector fields
shown in Fig.~\ref{fig:emradquad}. In
addition to ``stagnation points,'' the eigenlines contain line
singularities in the equatorial plane. In sequences of LIC snapshots
the motion of these singularities, and the understanding that 
they ``drag'' the field lines,  give approximate meaning to the 
evolution of the eigenlines.  Such qualitative visual 
flow fields could be used in making the movies, i.e., sequences
of LIC snapshots, that have been made available online\cite{onlineDLIC}.
However, we chose not to do that even qualitatively, and have taken
the underlying flow field to be zero in those movies.  The eye of the
observer provides a concept of motion in any case, and we felt that it
would be overinterpretation to try to augment that perception by
imposing a qualitative flow pattern that matches it.  Therefore we do
not evolve the underlying pattern at all in the gravitational
radiations movies online.

What may be the most interesting point about these visualizations of eigenline
dynamics is that, unlike motion of fluid velocity fields and electric field
lines, there is no ``correct'' and ``true'' motion of the eigenlines. There 
is neither a physical process (e.g., the spiraling of charged particles around 
magnetic field lines, the identification of specific fluid elements in velocity flows),
nor a mathematical criterion (e.g., flux conservation) available as a basis for defining
or constraining the motion of lines. Exploiting singularity location is the best we can
do. 

The intuitive feeling that there should be a well defined meaning to the motion
of field lines may be rooted in the relationship of electromagnetic field line 
motion and energy flow, a relationship, expressed in Eq.~(\ref{eq:vandPoynting}),
limited to regions, such as the radiation region, in which the {\bf E} and {\bf B} 
fields are orthogonal.  This suggests the question of whether energy flow could be 
used as a guide to the motion of eigenlines, and hence the question whether there 
is a gravitational analog of the Poynting flux. 

There is, in fact, a pragmatic quantitative measure of energy flow in
gravitational waves, the Landau-Lifschitz pseudotensor\cite{LL}, but
this measure is neither definitive nor useful for our purposes.
The fact that it is not definitive is important.
Just as gravitational acceleration near the Earth surface vanishes in
a freely falling elevator, many other aspects of gravitation vanish in
an appropriately chosen reference frame. A consequence of this is the
fundamental impossibility of localizing energy in a gravitational wave. The 
Landau-Lifschitz pseudotensor can give only a sort of average energy 
flux over several wavelengths. 

This already suggests that such a measure is not useful for us; it cannot help us give meaning to 
line motion. The mathematical details confirm that suggestion; the pseudotensor
cannot be inferred from the gravitoelectric and gravitomagntic fields. In a very rough
sense the pseuodensor is constructed from mathematical objects that are spacetime 
integrals of 
$\calEbf$ and
$\calBbf$\cite{handcurlies}.


\section{Conclusions}
\label{s:conc}

We have compared the mathematics and visualizations of electromagnetic
and gravitational fields, working up from static field configurations
to radiating oscillatory versions of these configurations. Most of the
details of the mathematics and the images were illustrated with the
examples of electric and gravitational point quadrupoles. Visualizations, using 
LICs, were based on the familiar field lines for the electromagnetic case.
Visualizations of gravitational fields used the recently introduced tendex/vortex eigenline formalism\cite{CorntechPRL,CorntechPRD,CorntechZeroes}.

We have found, as foretold in the Introduction, that this comparison
shows both instructive similarities and instructive differences. An
important similarity of radiation examples was the change, in both
cases, from a quasistatic field structure at distances from the point
source small compared to a wavelength, to a transverse, $1/r$,
radiation field at distances large compared to a wavelength.
The visualizations, in both cases, show the field structures
that are not easily seen in the mathematics, and show how the fields
make the transition from the  near-source structure to the 
very different radiation structure.

An important difference between the two cases lay in the visualization
of dynamical fields.  The motion of electromagnetic field lines has
both mathematical and physical meaning, while the motion of
gravitational eigenlines, has neither.  This difference can be
partially ascribed to the difference between the nature of eigenlines
for a tensor field and ``lines of force'' for a vector field. But the 
difference, especially regarding energy flow, underscores fundamental 
differences between electromagnetism and relativistic gravitation. 

In Sec.~\ref{sub:visdynaGrav} there was a rediscovery and illustration
of an interesting mathematical relationship between the
electromagnetic and gravitational fields.  Aside from trivial
replacements, the complete details of electromagnetic solution --
including the radiation fields -- is contained within the
gravitational field solution.  To go from the electromagnetic solution
to the gravitational, requires only adding the gravitational radiation
fields.


\begin{acknowledgments}
For discussions of tendex/vortex lines
we thank the groups at Caltech and Cornell, in  particular Kip Thorne, 
Mark Scheel, Rob Owen, Jeff Kaplan, Fan Zhang and Aaron Zimmerman.
\end{acknowledgments}


\begin{thebibliography}{5}
\bibitem{noSRT} We omit special relativistic effects here, such 
as the Lorentz factor. We assume that motions are slow compared to 
light speed $c$.
\bibitem{relversion} In fully nonlinear general relativity 
the gravitoelectric and gravitomagnetic tensor are defined 
as projections of the Weyl tensor (equal to the Riemann tensor
in source free spacetime). For details see Ref.~\onlinecite{CorntechPRD}.
\bibitem{bianchiident} In general relativity these equations follow
  from the the ``Bianchi identities,'' mathematical identities satisfied by
the
Riemann tensor. Since ${\bs{\cal B}}$ has no Newtonian equivalent, these
equations cannot be understood from simple Newtonian gravity theory.
\bibitem{c2factor} In literature on general relativity, units are
  often adopted in which $c$ becomes unity. As pointed out in
  Sec.~\ref{s:intro}, we do not make that choice here. In accordance
  with Eq.~(\ref{eq:geodev}), with our definitions ${\cal E}$ has
  dimensionality 1/time$^2$ and ${\cal B}$ has dimensions
  1/(time$\cdot$length).
\bibitem{vsmetricperts} We emphasize this point because weak relativistic gravitational
fields are often described with metric perturbations $h_{\mu\nu}$. These perturbations,
like electromagnetic potentials, are mathematically useful, but they are subject to 
gauge transformations and hence do not directly represent physical effects.
\bibitem{allpositive} This ellipsoid can be drawn only if the three principal
moments of inertia are all positive, but the fact that mass density is nonnegative
does guarantee this. The ellipsoid cannot be used so simply for, e.g., ${\cal E}_{jk}$
since its tracelessness means that the sum of its principal values is zero.
\bibitem{neg2M}This inclusion is made for simplicity
to eliminate the mass monopole when the $d\rightarrow0$ limit is taken.
With the $-2M$ particle omitted, the argument would require eliminating
the monopole moment by other means, such as a specific projection only 
of the quadrupole moment.
\bibitem{whynocalB} 
In the gravitational case this follows from the fundamental definition 
of ${\cal B}$ in general relativity. 
${\cal B}$ 
 involves the components of 
a fourth rank curvature tensor with only a
single time (i.e., 0) index. Such components reverse sign under time
reversal. In static configurations these components must be unchanged 
under time reversal, hence they must be zero. The vanishing of ${\cal B}$ for static configurations 
is also suggested by Eq.~(\ref{eq:curlyBeqs}).
\bibitem{EMcalcs}
These fields are most simply computed using the vector spherical 
harmonics in Sec.~9.10 of Ref.~\onlinecite{JacksonQ}
\bibitem{family}Each curve in 
Fig.~\ref{fig:LICGsnapshot} corresponds to a particular eigenvalue. Each family 
represents a set of eigenvalues and of eigenvector components that vary smoothly 
from one curve to the adjacent curve. It turns out that the signs of the eigenvalues
vary within a family. The eigenvalues for the top LIC in Fig.~\ref{fig:LICGsnapshot}
are negative both in the near zone and the radiation zone, but positive in a part 
of the intermediate zone. The opposite applies to the bottom LIC.
\bibitem{handcurlies} The Landau-Lifschitz pseudotensor is constructed
from the spacetime gradients of metric perturbations. The gravitoelectric
and gravitomagnetic fields are constructed from the gradients of the gradients
of the metric perturbations.
\end{thebibliography}
\end{document}